\documentstyle[epsf]{mn}

\newif\ifAMStwofonts

\def\simlt{\lower.5ex\hbox{$\; \buildrel < \over \sim \;$}}
\def\simgt{\lower.5ex\hbox{$\; \buildrel > \over \sim \;$}}

\ifoldfss
  \ifCUPmtlplainloaded \else
    \NewTextAlphabet{textbfit} {cmbxti10} {}
    \NewTextAlphabet{textbfss} {cmssbx10} {}
    \NewMathAlphabet{mathbfit} {cmbxti10} {} 
    \NewMathAlphabet{mathbfss} {cmssbx10} {} 
  \fi
  \ifAMStwofonts
    \ifCUPmtlplainloaded \else
      \NewSymbolFont{upmath} {eurm10}
      \NewSymbolFont{AMSa} {msam10}
      \NewMathSymbol{\upi}     {0}{upmath}{19}
      \NewMathSymbol{\umu}     {0}{upmath}{16}
      \NewMathSymbol{\upartial}{0}{upmath}{40}
      \NewMathSymbol{\leqslant}{3}{AMSa}{36}
      \NewMathSymbol{\geqslant}{3}{AMSa}{3E}

    \fi
  \fi
\fi 

\ifnfssone
  \newmathalphabet{\mathit}
  \addtoversion{normal}{\mathit}{cmr}{m}{it}
  \addtoversion{bold}{\mathit}{cmr}{bx}{it}
  \newmathalphabet{\mathbfit} 
  \addtoversion{normal}{\mathbfit}{cmr}{bx}{it}
  \addtoversion{bold}{\mathbfit}{cmr}{bx}{it}
  \newmathalphabet{\mathbfss} 
  \addtoversion{normal}{\mathbfss}{cmss}{bx}{n}
  \addtoversion{bold}{\mathbfss}{cmss}{bx}{n}
  \ifAMStwofonts
    \ifCUPmtlplainloaded \else
      \UseAMStwoboldmath
      \makeatletter
      \new@mathgroup\upmath@group
      \define@mathgroup\mv@normal\upmath@group{eur}{m}{n}
      \define@mathgroup\mv@bold\upmath@group{eur}{b}{n}
      \edef\UPM{\hexnumber\upmath@group}
      \new@mathgroup\amsa@group
      \define@mathgroup\mv@normal\amsa@group{msa}{m}{n}
      \define@mathgroup\mv@bold\amsa@group{msa}{m}{n}
      \edef\AMSa{\hexnumber\amsa@group}
      \makeatother
      \mathchardef\upi="0\UPM19
      \mathchardef\umu="0\UPM16
      \mathchardef\upartial="0\UPM40
      \mathchardef\leqslant="3\AMSa36
      \mathchardef\geqslant="3\AMSa3E
    \fi
  \fi
\fi 

\ifnfsstwo
  \DeclareMathAlphabet{\mathbfit}{OT1}{cmr}{bx}{it}
  \SetMathAlphabet\mathbfit{bold}{OT1}{cmr}{bx}{it}
  \DeclareMathAlphabet{\mathbfss}{OT1}{cmss}{bx}{n}
  \SetMathAlphabet\mathbfss{bold}{OT1}{cmss}{bx}{n}
  \ifAMStwofonts
    \ifCUPmtlplainloaded \else
      \DeclareSymbolFont{UPM}{U}{eur}{m}{n}
      \SetSymbolFont{UPM}{bold}{U}{eur}{b}{n}
      \DeclareSymbolFont{AMSa}{U}{msa}{m}{n}
      \DeclareMathSymbol{\upi}{0}{UPM}{"19}
      \DeclareMathSymbol{\umu}{0}{UPM}{"16}
      \DeclareMathSymbol{\upartial}{0}{UPM}{"40}
      \DeclareMathSymbol{\leqslant}{3}{AMSa}{"36}
      \DeclareMathSymbol{\geqslant}{3}{AMSa}{"3E}
    \fi
  \fi
\fi 

\ifCUPmtlplainloaded \else
  \ifAMStwofonts \else 
    \def\upi{\pi}
    \def\umu{\mu}
    \def\upartial{\partial}
  \fi
\fi

\title[The inflaton potential and flow-equations]
	{Observational constraints on the inflaton potential
combined with flow-equations in inflaton space}
\author[Hansen \& Kunz]
{Steen H. Hansen, Martin Kunz\\
Department of Physics, Nuclear \& Astrophysics Laboratory,
University of Oxford, Keble Road, Oxford OX1 3RH, U.K.}
\date{Draft version \today}
\pagerange{\pageref{firstpage}--\pageref{lastpage}}
\pubyear{2002}

\begin{document}

\maketitle

\label{firstpage}

\begin{abstract}
Direct observations provide constraints on the first two derivatives
of the inflaton potential in slow roll models. We discuss how present
day observations, combined with the flow equations in slow roll
parameter space, provide a non-trivial constraint on the third
derivative of the inflaton potential. We find a lower bound on the
third derivative of the inflaton potential $V'''/V > -0.2$. We also
show that unless the third derivative of the inflaton potential is
unreasonably large, then one predicts the tensor to scalar ratio, $r$,
to be bounded from below $r > 3\times 10^{-6}$.
\end{abstract}



\section{Introduction}
Inflation is today considered a natural and necessary part of the
cosmological standard model, providing the initial conditions for
cosmic microwave background radiation and large scale structure
formation. Our know\-ledge of the fundamental physics responsible for
inflation is, however, very limited, and only recent observations of
the cosmic microwave background~\cite{boom,maxima,dasi} and large
scale structure~\cite{croft,pscz,efstathiou} have provided the first
glimpse of the underlying physics. This has been achieved (and is
still only possible) in slow roll inflation (see \cite{Lyth:1999xn}
for a review on slow roll and list of references).

For any given inflationary model one can find the power spectrum of
primordial curvature perturbations, ${\cal P}(k)$, which is a function
of the wavenumber $k$.  This power spectrum can be Taylor-expanded
about some wavenumber $k_0$ and truncated after a few
terms~\cite{Lidsey:1997np}
\begin{eqnarray}
 {\rm ln} {\cal P}(k)
&=& {\rm ln} {\cal P}(k_0) + (n_S-1) \, {\rm ln} \frac{k}{k_0}  \nonumber \\
&&+\left. \frac{1}{2} \frac{d\, n_S}{d \, {\rm ln}k} \right|_{k_0} \, {\rm
  ln} ^2 \frac{k}{k_0} + \cdots
\label{power}
\end{eqnarray} 
where the first term is a normalization constant, the second is the
power-law approximation, with the case $n_S = 1$ corresponding to a
scale invariant (Harrison-Zel'dovich) spectrum, and the third term is
the running of the spectral index.

Early data analyses~\cite{wang,al} have truncated this expansion after
the first two terms, hence assuming that the bend of the spectrum is
zero, $\partial_{{\rm ln} k} \equiv dn_S/d{\rm ln}k|_{k=k_0}=0$.
However, as shown in~\cite{Copeland:1997mn,Hannestad:2001tj}, 
this early truncation
gives too strong constraints on both scalar and tensor
indices, and the analysis must allow for a bend of the spectrum. In
most slow roll (SR) models $\partial_{{\rm ln} k}$ is expected to be
very small, since it is second order in small
parameters~\cite{Kosowsky:1995aa}, but there are very interesting
models where this need not be the
case~\cite{stewart,stewart2,Kinney:1998dv,DodStew}, and $\partial_{{\rm ln} k}$
may assume values big enough to be observable~(see
e.g. refs.~\cite{Copeland:1997mn,Covi:1999mb}).  The more general SR
models are constrained through the expansion~(\ref{power}), which can
provide constraints on the first two derivatives of the inflaton
potential~\cite{liddleturner,Hannestad:2001nu}.

In SR it is straight forward to find the derivatives of the scalar and
tensor spectral indices~\cite{Kosowsky:1995aa,Liddle:1992wi}, $d n_S /
d {\rm ln} k$ and $d n_T / d {\rm ln} k$, and these two provide the
flow equations in SR space~\cite{hoffman}.  We discuss below how one
can combine present day observation with these flow equations to
obtain a non-trivial bound on the third derivative of the inflaton
potential, $V'''/V$ (or combinations like $V'V'''/V^2$), under the
assumption that $V'''/V$ (or $V'V'''/V^2$) can be treated as
approximately constant.

\section{Slow roll models}
{\bf The flow equations}.
Slow roll models are traditionally defined through the 3 parameters
$\epsilon,\eta$ and $\xi^2$, which roughly correspond to the first,
second and third derivatives of the inflaton potential. We will use
the notation~\cite{Lyth:1999xn}
\begin{eqnarray}
\epsilon \equiv \frac{M^2}{2} \left( \frac{V'}{V} \right)^2 
\, \, \,  ,  \, \, \, 
\eta \equiv M^2 \frac{V''}{V} \, \, \, , \, \, \,
\xi^2 \equiv M^4 \frac{V' V'''}{V^2}  \, ,
\label{definitions}
\end{eqnarray}
where $M$ is the reduced Planck mass, $M=2.4 \times 10^{18}$ GeV,
from which one can express the SR parameters using the directly
observable quantities $n,r $ and $\partial_{{\rm ln} k}$
\begin{eqnarray}
2 \xi^2 &=& -\partial_{{\rm ln} k} - 24 \epsilon^2 +16 \epsilon \eta \, , 
\label{eqsxi} \\
2 \eta &=&n_S-1 + 6 \epsilon \, ,  \label{eqseta}\\
2 \epsilon &=& \frac{r}{\kappa} = - n_T \label{epslignt} \, ,
\end{eqnarray}
where $r$ is the tensor to scalar ratio at the quadrupole. 
Eqs.~(\ref{eqseta},\ref{epslignt}) are truncated at order $\xi^2$
and eq.~(\ref{eqsxi}) at order $V'^2V''''/V^3$, and are thus
correct to leading order in slow roll expansion.

The factor $\kappa$ in eq.~(\ref{epslignt}) depends on the given
cosmology~\cite{knox,Turner:1996ge}, in particular on the value of
$\Omega_\Lambda$ and $\Omega_M$, and in this paper we will use the
value $\kappa = 5$, corresponding to $\Omega_\Lambda=0.65$ and
$\Omega_M = 0.35$.

As the inflaton rolls down the potential, the values of $n_S$ and $n_T$
will change, and this variation is governed by the flow
equations~\cite{Liddle:1992wi,Kosowsky:1995aa}
\begin{eqnarray}
\frac{d \, n_S}{d \, N} &=& - 4 \frac{r}{\kappa} \left[
\left(n_S-1 \right) + \frac{3}{2} \frac{r}{\kappa}
\right] + 2 \xi^2 \, ,
\label{flow1} \\
\frac{d \, n_T}{d \, N} &=& -\frac{r}{\kappa} \left[
\left(n_S-1 \right) + \frac{r}{\kappa} \right] \, ,
\label{flow2}
\end{eqnarray}
where we have used $d \, {\rm ln} k = -dN$ with $N$ the number of
Hubble times (e-folds) until the end of inflation.
Also this equation is correct at leading order in slow roll, since
one has $d/dN = -(1-\epsilon)d/d{\rm ln}k$~\cite{liddleturner}. 
The connection between $\xi^2$ and
$M^3 V'''/V$ through $\epsilon$ is given by 
equations~(\ref{definitions}), 
$\xi^2 = \sqrt{r/\kappa} M^3 V'''/V$.
Certainly one can find good 
inflationary models, which do not obey this slow-roll description. This could
e.g. appear, because the derivation of the slow-roll equations
is based on the assumption of a slowly varying Hubble parameter, which
for particular models could be violated.

\begin{figure}
\begin{center}
\epsfxsize=7.2cm
\epsffile{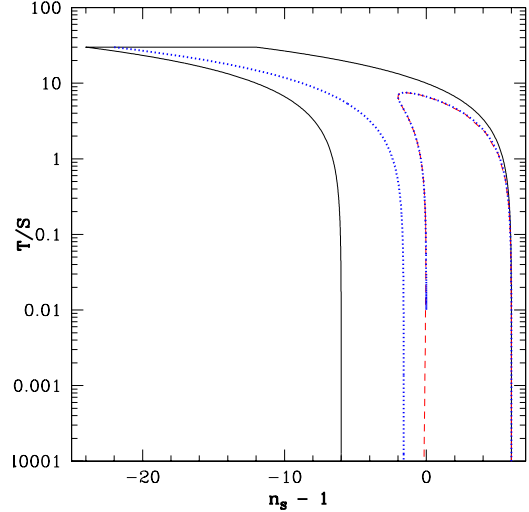}
\end{center}
\caption{The SR validity region (solid black line) flown
back 50 e-folds (dashed red line) together with two examples
of flow lines (blue dashed lines) 
ending on the boundary of the SR validity region. 
In this case we set $\xi^2 = 0$.}
\label{fig1}
\end{figure}

As $N$ decreases, the inflaton rolls down its potential, and the
observable parameters are determined when the relevant scales cross
outside the horizon, approximately 50-60 e-folds before the end of
inflation~\cite{kolbturner}.  Single field inflation 
will end, when the SR conditions
are violated~\cite{kolbturner,hoffman}
\begin{equation}
r < 6\kappa \, \, \, \, \, \mbox{or} \, \, \, \, \, 
\left| \left( n_S-1 \right) + \frac{3}{\kappa} r \right| < 6 \, .
\label{boundary}
\end{equation}
The area in $(n_S,r)$ space inside this boundary is denoted the SR
``validity-region''. The solid lines in
fig.~\ref{fig1} show this region, and also examples of the flow
of two models (dotted lines).
An almost trivial observation is that the
region allowed by current observations, given in (\ref{obsns}-\ref{obsdlnk}), 
lies inside the SR validity-region.

In order to close the set of equations, so that the flow equations
uniquely define the time evolution of $n_S$ and $n_T$, we need to
introduce an additional constraint, and there are various
possibilities.  In ref.~\cite{hoffman} the assumption was made that
$x''=0$, where $x=V'/V$.  Another possibility would be to assume that
either $V'''/V$ or $\xi^2$ can be treated as constants. Different
choices will lead to different fix-points and different time evolution
of $n_S$ and $n_T$, and general conclusions are therefore only
credible if such conclusions are reached for any choice of this
additional constraint.


{\bf Observational constraints}.  The COBE
observations~\cite{Bunn:1996py} gave the first constraint on the first
derivative
\begin{equation}
\frac{V^{3/2}}{M^3 V'} \approx 5 \times 10^{-4} \, ,
\end{equation}
and the present day constraints on slow roll parameters are improved
when combining CMB data with data from the Lyman-$\alpha$ forest. The
reason being that the error-ellipses for CMB and Lyman-$\alpha$ are
almost perpendicular~\cite{Hannestad:2001nu}.  The reason for using
Lyman-$\alpha$ data~\cite{croft} instead of ``standard'' LSS data
(such as PSCz~\cite{pscz} or 2dFGRS~\cite{efstathiou}) is, that the
Lyman-$\alpha$ data are obtained at high red-shift, where small scales
are still linear.  One should naturally keep in mind, that neither CMB
nor Ly-$\alpha$ data include all the possible systematic errors.  The
bounds obtained are~\cite{Hannestad:2001nu} (all at $2\sigma$)
\begin{eqnarray}
0.8 < & n_s & < 1.0 \, , \label{obsns} \\
 0 < & r & < 0.3 \, ,  \label{obsr} \\
-0.05 < & \partial_{{\rm ln} k} & < 0.02 \, , \label{obsdlnk}
\end{eqnarray}
where $n_S$ is the scalar spectral index, $r$ is the tensor to scalar
ratio, and $\partial_{{\rm ln} k} = dn_S/d{\rm ln}k$ is the bend
defined through eq.~(\ref{power}).  These bounds directly provide
constraints on the first and second derivatives of the potential,
\begin{eqnarray}
M \left| \frac{V'}{V} \right| & < & 0.25 \, ,  \label{obsv'} \\
M^2 \left| \frac{V'' }{V} \right| & < & 0.1 \, , \label{obsv''} 
\end{eqnarray}
however, the third derivative is not directly constrained. 
Instead, eqs.~(\ref{obsns}-\ref{obsdlnk}) limit only $\xi^2$ to be smaller
than about $|\xi^2| < 0.036$, when assuming independent errors on $n_S,r$ and 
$dn_S/d{\rm ln}k$, and in reality one could obtain a slightly stronger bound.
To obtain a
bound on $V'''$ we must combine the observational constraints
(\ref{obsns}-\ref{obsdlnk}) with the flow equations in SR parameter
space, eqs.~(\ref{flow1},\ref{flow2}). This is because one has
$dn_S/d{\rm ln}k \approx \sqrt{r} \, (V'''/V) \, (-2M^3/\sqrt{\kappa}) 
+ 4r/\kappa\,
[n_S-1 + 3/2 \, r/\kappa]$, and since we don't have a lower
bound on $r$ we cannot get any direct constraint on $V'''/V$.

\section{Discussion}
We are going to consider the case, where slow-roll inflation is ended
because the slow-roll conditions are violated, eq.~(\ref{boundary}). 
Another possibility
would be to allow for other fields coupled to the inflaton field which
could end inflation. The parameters observable with CMB and LSS are
determined approximately 50 e-folds before the end of inflation, and
we therefore run the SR violating boundary back in time 50
e-folds.  
This is done for
various values of fixed $\xi^2$ (or fixed $V'''/V$).  Now we demand
that the observable parameters be in agreement with eq.~(\ref{obsns}),
and if no point on the SR violating boundary lands inside the observed
parameter-range, then we can exclude this value of $\xi^2$ (or
$V'''/V$).

\begin{figure}
\begin{center}
\epsfxsize=7.2cm
\epsffile{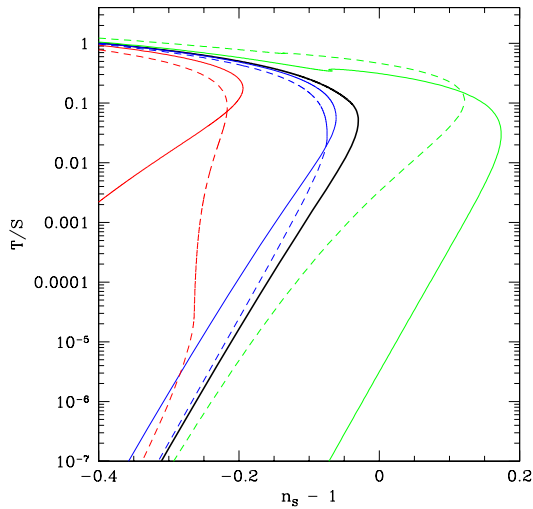}
\end{center}
\caption{Examples of the flown back SR validity boundary, 50
e-folds before the end of inflation. The thicker solid black
line shows the case for $\xi^2 = 0$. For the solid lines, 
$\xi^2$ was kept constant, while the dashed lines are for
models where $V'''/V$ was kept constant. From left to right
we show examples for $\xi^2 = -10^{-2}$, $-10^{-3}$ and
$+5\times 10^{-3}$ as well as $V'''/V = -5\times 10^{-2}$,
$-10^{-2}$ and $+5\times 10^{-2}$. For small $r = T/S$ the models
with constant $\xi^2$ move nearly parallel to the $n_S$
direction while the ones with constant $V'''/V$ are frozen and
remain close to the $\xi^2=0$ curve.}
\label{fig2}
\end{figure}

Let us first consider the case where $\xi^2$ can be treated as a
constant during the 50 e-folds. If $\xi^2=0$ one finds, that 50 e-folds
before the crossing of the SR violating boundary one has
\begin{equation}
2 \times 10^{-5} < r < 0.5 \, ,
\end{equation}
when we demand that $n_s$ complies with eq.~(\ref{obsns}). This can also
be seen in fig.~\ref{fig2}, where the thicker solid line is for
$\xi^2=0$.  If $\xi^2$ is positive then $r$
must be even smaller, since the region in fig.~\ref{fig2} moves to the right
(larger $n_s$), and for $\xi^2 > 0.06$ there are no more points in
agreement with observations.  For negative $\xi^2$ the acceptable
values of $r$ are larger than $2\times 10^{-5}$, and for $\xi^2 <
-0.06$ there are no points in agreement with observations. We hence
conclude, that in the case where $\xi^2$ can be considered constant
throughout the 50 e-folds and inflation ends by violating the
slow-roll conditions, one must have $|\xi^2| < 0.06$. We therefore find 
a constraint on $\partial_{{\rm ln} k} $ similar to (\ref{obsdlnk}) 
from the observational constraints on $n_s$ and the flow equations alone.
This bound on $\xi^2$ can be converted into a bound on $V'''/V$ using
the predicted $r$ and the relation $\xi^2 = \sqrt{r/\kappa} M^3 V'''/V$.
We find $M^3 V'''/V > -0.2$.

If one instead considers the case where $V'''/V$ can be treated as a
constant during the 50 e-folds, then the conclusions are slightly
different. Again, in the case with $V'''/V=0$ one finds
$2 \times 10^{-5} < r < 0.5$. When $V'''/V$ is negative then no
points agree with observations if $V'''/V < -0.05$. Only for positive
$V'''/V$ there will always be acceptable points, however, the allowed 
range for $r$ will decrease, e.g. for $M^3 V'''/V =1$ one finds 
$10^{-7} < r <10^{-5}$.

As discussed above, the most credible results must agree independently
of the additional constraint (fixed $\xi^2$ or fixed $V'''/V$). We have
seen that one always finds a lower bound
\begin{equation}
\frac{V'''}{V} > - 0.2\, .
\end{equation}
This is the first constraint found on the third derivative of the 
inflaton potential, and is valid under the assumptions specified 
above.
It is important to
note, that the approach adopted here differs from the results
of ref.~\cite{liddleturner}, where it was pointed out, that an
observation of $\partial_{{\rm ln} k}$ would provide knowledge about
$V'''$.  The difference being, that in our case we have only an
observational {\em upper bound} on $r$, and without the use of the
flow equations, this will leave $V'''$ completely unknown.

No strong predictions can be made on the magnitude of $r$, simply
because if $V'''/V$ is large then $r$ is allowed to be smaller,
however, one would often expect $M^3 V'''/V$ to be smaller than both
$M^2 V''/V$ and $M V'/V$, eqs.~(\ref{obsv'},\ref{obsv''}), in which
case one predicts $r > 3\times 10^{-6}$.

The number of e-folds $N$ depends on the detailed mechanism of
inflation such as the reheat temperature and the energy scale
of inflation, and can be somewhat
different from 50 (see e.g. \cite{Lyth:1999xn}). If $N=60$ our
direct bound on $V'''/V > - 0.05$ remains unchanged, however,
the inferred bound from the case of constant $\xi^2$ is weakened
by approximately a factor of 2. The lower bound on r discussed
above becomes $r > 3\times 10^{-7}$. Naturally, for a lower value
of $N$ the bounds are correspondingly stronger.

\section{Conclusion}
COBE gave us the first clear information on the first derivative of
the inflaton potential, and the combination of CMB observations with
data from the Lyman-$\alpha$ forest has given us information on the
first two derivatives of the inflaton potential. Here we have combined
the present observations with the flow equations in slow roll space,
and found a lower bound on the third derivative of the inflaton potential 
$V'''/V > -0.2$. We have also shown, that unless $V'''/V$ is 
unreasonably large, then one predicts the tensor to scalar ratio, $r$,
to be bounded from below  $3\times 10^{-6} < r $.

\section*{Acknowledgements}
It is a pleasure to thank Pedro Ferreira and Francesco Villante for
comments and discussions, and Massimo Hansen for inspiration. SHH is
supported by a Marie Curie Fellowship of the European Community under
the contract HPMFCT-2000-00607. MK is supported by a Marie Curie
Fellowship of the Swiss National Science Foundation under the contract
83EU-062445.

\label{lastpage}

\end{document}